\documentstyle[aps,epsf]{revtex} 
\begin{document}   
\draft   
\title{Dislocation loops in overheated free-standing smectic films}
\author{A. N. Shalaginov and D. E. Sullivan }   
\address{Department of Physics and Guelph-Waterloo Physics 
Institute, University of Guelph, Guelph,  
Ontario N1G 2W1, Canada }   
\date{\today}   
\maketitle   
\begin{abstract}
Static and dynamic phenomena in overheated free-standing smectic-A 
films are studied theoretically. The work is based on a generalization,
introduced recently by the authors,
of de Gennes' theory for a confined presmectic liquid.
In this approach, smectic ordering in an overheated film
is caused by an intrinsic surface contribution to the film free energy
and vanishes at some temperature depending on the
number of layers. Here the theory is further generalized to study the dynamics
of films with planar inhomogeneities. A static application is to determine
the profile of the film meniscus and the meniscus contact angle, the
results being compared with those of a recent study employing
de Gennes' original theory. The dynamical generalization of the theory
is based on a time-dependent Ginzburg-Landau approach. This
is used to compare two modes for layer-thinning transitions
in overheated free-standing films, namely ``uniform thinning'' {\it vs.}
nucleation of dislocation loops. It is concluded that the nucleation
mechanism dominates provided there is a sufficiently large pressure
difference arising from meniscus curvature. Properties such as
the line tension and velocity of a moving dislocation line are
evaluated self-consistently by the theory. 
\end{abstract}   
%
%

\section{INTRODUCTION}\label{sec:non_intro}

Free-standing films of several smectic-A liquid-crystalline compounds
can be heated above the bulk smectic disordering temperature without
immediately rupturing, and instead are found to undergo successive 
layer-by-layer thinning transitions as the temperature is increased
\cite{stoprl94,joh97,pan98,dem95rc,mol98,clu00}.
The persistence of smectic layering in an overheated thin film
is usually attributed to
enhanced ordering associated with the free surfaces of the film,
as is known to occur in other contexts \cite{bahr94b}.  There is not yet, however, a
clear consensus on the mechanisms by which layer thinning occurs. 
According to one set of theories \cite{mir96,mar97,gor99,sha01}, 
thinning takes place when the smectic layer
structure in the middle of a film vanishes. In an
alternative theory \cite{pan99}, supported by experimental studies
\cite{pan00}, layer thinning occurs by spontaneous nucleation of dislocation
loops prior to the melting of the layer structure in the film interior. This
mechanism is not necessarily unrelated to the former, since a sufficient
reduction in the degree of interior smectic ordering is required for it to
proceed.

One of the key experimental
observables is the variation of layer-thinning transition temperatures
$T_c(N)$ with the number of film layers $N$, which is found to be well
fit by the power-law relation $N \propto t^{-\nu}$, where $t =
(T_c(N)-T_0)/T_0$, $\nu \approx 0.70 \pm 0.10$, and $T_0$ is close to
the bulk transition temperature.
Alternative mathematical relations \cite{gor99,pan99,pic01} and an
upper bound \cite{sha01} for $T_c(N)$ have been derived from the
different theories.
With appropriate fitting parameters, these alternative relations
all turn out to agree well with the power-law
expression and, thus, are not able to distinguish between
the various mechanisms.

In this paper, we examine further the connections between the
different proposed mechanisms of layer-thinning transitions.
As in several recent works \cite{gor99,pan99,pic01}, our analysis is based on de
Gennes' \cite{gen90} phenomenological Landau theory for a ``presmectic'' film
of a fluid exhibiting a {\it second-order} bulk A-nematic (N) transition.  More precisely, we employ a
generalization of that theory recently proposed by the present authors
\cite{sha01}.  One drawback of de Gennes' original model 
stems from attributing surface-enhanced smectic ordering to an
external-field-like coupling term of constant magnitude, which is more
appropriate for a film confined between solid walls.  This has the
consequence that a weak degree of smectic ordering in a thin film is predicted
to persist up to arbitrarily high temperatures.  In order to induce layer
thinning, recent studies based on this theory have included the effects of
a pressure difference $\Delta P$ associated with curvature of the meniscus at
the film border \cite{gor99,pic01}. According to the latter studies, $\Delta
P$ must be of a sufficient magnitude to cause layer thinning, although the
layer-thinning transition temperatures are found to depend only
logarithmically on $\Delta P$. The modified version of de Gennes' theory
proposed by the present authors \cite{sha01} utilizes a different form of the
surface contribution to the free energy (suggested by older theories of
wetting \cite{sul86}), which is quadratic rather than linear in the surface
order parameter and which restores smectic melting at high temperatures 
without requiring the pressure term.

According to the theory of Ref.\cite{sha01}, reviewed here in
Sec.~\ref{sec:non_unif}, the free energy per area of an overheated smectic
film exhibits a discrete sequence of metastable local minima, whose depths
decrease with increasing number of layers up to some finite maximum $N_{cr}$
depending on temperature.
At or slightly below the temperature for which the free-energy well at $N_{cr}$
vanishes, it was assumed in
Ref.\cite{sha01} that the film would spontaneously thin down to a smaller
thickness. This assumption makes no statement on ``how'' 
layer thinning occurs, which is the question addressed in this paper.
We proceed by generalizing the theory to allow for planar
inhomogeneities in the film (Sec.~\ref{sec:non_non}) as well as the dynamical
evolution of the film, using a time-dependent Ginzburg-Landau approach
(Sec.~\ref{sec:dynam}). In the static limit, the theory provides a description
of the contact angle between the film and meniscus, which we compare with the
recent experimental and theoretical studies of Picano et al. \cite{pic01}.
Using the dynamical theory, we investigate the nucleation of dislocation loops
between film domains of different thicknesses and the subsequent growth of the
thinner region of the film, which is contrasted with the ``uniform-thinning''
mechanism for the film to achieve a state of lower free energy.  The results
are mapped onto the conventional nucleation picture, in which the activation
free-energy barrier to nucleation depends on both the difference in
well-depths of the homogeneous film regions and the line tension $E$ of the
dislocation loop.  Here we calculate the line tension $E$, activation energy,
and velocity of a growing loop self-consistently in the dynamical model, and
evaluate their dependence on temperature and number of layers. 
It is found that thinning via nucleation of dislocation loops preempts 
the uniform-thinning mechanism 
provided the pressure difference $\Delta P$ due to the meniscus is
sufficiently large. Conclusions of our findings, and the fact that they are
restricted to smectic systems with continuous as opposed to first-order bulk transitions,
are discussed in  
Sec.~\ref{sec:disc}.

\section{Theory of uniform films  }\label{sec:non_unif}

In this section we review the generalized de Gennes theory in the case of a
uniform planar free-standing smectic film: further details can be found in
Ref.\cite{sha01}. The film is modelled by a thin liquid slab bounded by two
parallel surfaces located at $z=\pm L/2$, where $L$ is the film thickness. The
degree of smectic order in the film is represented by the complex order
parameter $\Psi(z)$, where the real part of $\Psi(z)\exp(i q_0 z)$ describes
spatial modulation of the density. Here $q_0 = 2\pi/d$, with $d$ being
the unstressed smectic layer spacing.  The order parameter is parameterized as
\begin{equation}
\label{Psi}
\Psi(z) =\psi(z)\exp[-i \phi(z)],
\end{equation}
where $\psi(z)$ is the amplitude and $\phi(z) \equiv q_0u(z)$ is a phase
proportional to the layer displacement $u(z)$.

The Landau free energy per unit area of the film \cite{foot1} is taken to be
\begin{eqnarray}
\label{f[eta,u]}
\nonumber
&f& = \frac{1}{2}\int\limits_{-L/2}^{L/2} dz \left[r\psi^2+\frac{1}{2}g\psi^4
+C (\nabla_z\psi)^2
+C \psi^2(\nabla_z \phi)^2 \right]\\ 
&& +\frac{1}{2}r_s\left[\psi^2(L/2)+\psi^2(-L/2)\right]
-h_s\left[\psi(L/2)+\psi(-L/2)\right]~,
\end{eqnarray}
where $C$ is an elastic constant. This model free energy generalizes that used
in Refs.~\cite{gor99,pan99,pic01,gen90} by including both a quartic term
$(g/4)\psi^4$ with $g > 0$ in the bulk free-energy density and quadratic
contact terms $r_s \psi^2(\pm L/2)/2$ \cite{sul86} in the contribution of the surface
layers. As in those works, the bulk free-energy density employed in
Eq.~(\ref{f[eta,u]}) is strictly applicable only to a system exhibiting a
second-order bulk smectic disordering transition, which occurs at $r=0$ in
this mean-field theory. We express $r=a(T-T^*)$, where $T^*$ denotes the bulk
mean-field transition temperature \cite{cha95}.  Euler-Lagrange equations
determining $\psi(z)$ and $\phi(z)$ are obtained by functional minimization of
Eq.~(\ref{f[eta,u]}).

Two alternative versions of this theory for surface-enhanced ordering of
smectic free-standing films have been investigated. In the first approach
\cite{gor99,pan99,pic01}, following the original work of de Gennes
\cite{gen90}, such ordering is induced solely by the surface field $h_s$, with $r_s
= 0$. These studies have also set $g = 0$. Refining some steps of these earlier
analyses, which were restricted to the asymptotic limit $L \gg \xi$, we find
that the equilibrium free energy of this model is given by \cite{sha01a}
\begin{equation}
  \label{eq:f_un}
f(L)=-h_s\psi(L/2)=-\frac{h_s^2\xi}{C}
\frac{\sinh(L/\xi)}{[\cosh(L/\xi)-\cos(q_0L)]}~,
\end{equation}
where $\xi \equiv \sqrt{C/r}$ is the bulk correlation length.  Note that
Eq.~(\ref{eq:f_un}) diverges on approaching the bulk critical temperature, as
does the order parameter $\psi(z)$ throughout the film, a consequence of
neglecting the non-harmonic term $g\psi^4$ in Eq.~(\ref{f[eta,u]}). This feature has been mitigated in
some versions of the theory \cite{gor99,pic01,mor94} by the somewhat {\it
  ad-hoc} procedure of fixing $\psi(L/2)$ instead of $h_s$. It is also
worthwhile to note that the minima of Eq.~(\ref{eq:f_un}) do not occur
precisely at the values $L = Nd$, but at slightly compressed values $L < Nd$
of the film thickness, which is due to an interplay between finite-size and
surface-ordering effects. This feature also holds for the more general model
in Eq.~(\ref{f[eta,u]}).
 
\begin{figure}
\centerline{\epsfxsize=400pt\epsffile{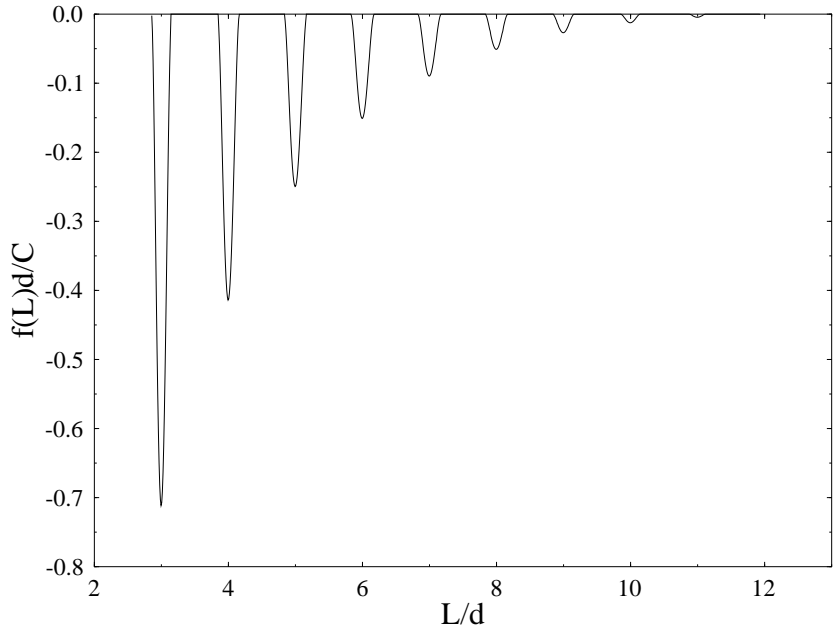}}  
\caption{ 
\label{fig:f_vs_l_r_s} 
The dimensionless free energy per unit area $fd/C$  {\it vs.} thickness
$L/d$ calculated using scaled parameters $h_s=0$,
$r_sd/C=-0.2$, $gd^2/C=0.01$ and $rd^2/C=0.05$.  }
\end{figure}

In Ref. \cite{sha01}, smectic ordering in an overheated film is attributed to
a non-zero value $r_s < 0$, with $h_s = 0$. In the following we refer to this 
as the $r_s$-model. In this case, the Euler-Lagrange equations obtained from
Eq.~(\ref{f[eta,u]}) always admit a trivial solution $\psi(z)=0$ describing a
disordered state of the film, which is the only solution at sufficiently high
temperature.  The numerical solution of those equations, which consider $g
\neq 0$, is described in Ref. \cite{sha01}. The variation of the equilibrium
free energy $f(L)$ typical for this model is depicted in
Fig.~\ref{fig:f_vs_l_r_s}. The free energy exhibits a set of wells with
centers situated approximately at $L=Nd$ and depths diminishing with $N$,
while $f(L)$ vanishes over non-zero ranges of $L$ between the wells. Although
not discernable in the figure, the slope of the free energy smoothly approaches
zero at the limits where $f(L) \rightarrow 0$.  In the temperature range $T <
T_s$, where $T_s = T^*+r_s^2/(Ca)$, $f(Nd)$ for any $N$ is less than some
negative threshold depending on $T$. For temperatures $T>T_s$, wells with
non-zero depths occur only up to a finite number $N_{cr}$.  This in the case
in Fig.~\ref{fig:f_vs_l_r_s}, where $N_{cr}=11$. It was argued in
Ref.~\cite{sha01} that the temperature at which the free-energy well for
$L=N_{cr}d$ disappears, which we will call the ``maximum 
temperature'' for an $N_{cr}$-layer film, is an upper limit for the layer-thinning transition
temperature $T_c(N_{cr})$. Films of all $N<N_{cr}$ can still exist as
metastable states and, in principle, thinning could then occur to any one of
these states.

We note the following scaling of the $r_s$-model free energy, used in
Fig.~\ref{fig:f_vs_l_r_s} and in subsequent analysis. On expressing distances
in units of the layer-spacing $d$, one sees that the free energy in Eq.~(\ref{f[eta,u]}) can be
expressed as $f(L,r,g,r_s,C) = (C/d)\hat f(L/d,\hat r,\hat g, \hat r_s)$,
where $\hat r = rd^2/C$, $\hat g = gd^2/C$, and $\hat r_s = r_sd/C$ . In
principle, the parameter $\hat g$ could also be pulled out from $\hat f$ by
suitable scaling of the order parameter $\psi$, but we have found it
convenient for numerical analysis to set this at the (arbitrary) small value
$\hat g = 0.01$ and leave the scaling of $f$ in the form indicated.

\section{Nonuniform films: statics}\label{sec:non_non}
In order to study dislocation loops and associated phenomena in free-standing
films, we need to generalize the theory of Sec.~\ref{sec:non_unif} to
incorporate planar inhomogeneities in a film. This could be
carried out by generalizing the free energy of Eq.~(\ref{f[eta,u]}) to include
horizontal gradients of the order parameter $\Psi$ and nematic director,
consistent with de Gennes' more general theory of the N-A transition
\cite{gen93}, as well as of the film thickness $L$. Full analysis of this
theory would require minimizing the resulting free energy with respect to
$\Psi$ and the director. In the remainder of this paper we will examine a
simplified phenomenological description using an interface-displacement theory
\cite{fis89}, in which the free energy is expanded in gradients of $L$. To
leading order in these gradients, the film free energy $F$ is given by
 \begin{equation}
  \label{eq:F_approx}
F=\int d^2r_\perp \left[f(L)+\frac{1}{2}D(\nabla_\perp L)^2\right].
\end{equation}
Here the horizontal direction is represented by ${\bf r}_\perp$, with
Cartesian components $(x,y)$, $\nabla_\perp=(\nabla_x,\nabla_y)$, $L=L({\bf
  r}_\perp)$ is the spatially-varying film thickness, and $f(L)$ is the
equilibrium free energy per area of a film of {\it uniform} thickness $L$,
obtained from Eq.~(\ref{f[eta,u]}). The coefficient $D$ characterizes the
``stiffness'' \cite{fis89} of the film surfaces, which in principle could be
derived by analysis of the underlying de Gennes theory. We expect that $D
\approx \gamma/2$, where $\gamma$ is the liquid-vapor interfacial tension. 
Generally $D$ could vary with $L$ due to changes in the degree of smectic
ordering, but such changes are expected to be weak and here we will assume
that $D$ is constant.

An additional term may enter Eq.~(\ref{eq:F_approx}) due to the existence of a
positive pressure difference $\Delta P = P_{air}-P_{liquid}$ across the
surface of the meniscus surrounding the film. Such a pressure difference
produces a shift $\Delta \mu$ in the chemical potential of the film molecules
from their value at coexistence with the vapor phase across a planar interface
\cite{tos75,pie93}. This leads to the replacement of $f(L)$ in
Eq.~(\ref{eq:F_approx}) by $\tilde f(L)$, where
\begin{equation}
  \label{eq:tilde_f_def}
  \tilde f(L)=f(L)+\Delta PL. 
\end{equation}
The main effect of the $\Delta P$ term is to shift the depths and to a slight extent 
the positions of
the smectic minima of the effective free energy $\tilde f(L)$ with respect to
those of $f(L)$, possibly eliminating minima occurring at large $L$
\cite{gor99}.

The static free energy Eq.~(\ref{eq:F_approx}) can be used to obtain the shape
of the meniscus at the edge of a free-standing film, following the analysis
of Ref.~\cite{pic01} based on the original de Gennes model.  Related
considerations have been applied to describe the shape of liquid droplets on a
solid substrate \cite{saf94,pis00}.  The profile of the meniscus is found by
minimizing Eq.~(\ref{eq:F_approx}) with respect to $L$ (replacing $f$ by
$\tilde f$), subject to the boundary condition that the main part of the film
has a thickness $L \equiv H \approx Nd$.  Assuming that $L$ varies only in the
$x$-direction, the resulting Euler-Lagrange equation is
\begin{equation}
  \label{eq:L_x_equation}
  D\frac{d^2 L}{dx^2}-\frac{\partial \tilde f}{\partial L}=0~.
\end{equation}
The first integral of this equation, with the boundary condition
$L(-\infty)=H$, is
\begin{equation}
\label{eq:first_integral}
\frac{D}{2}\left(\frac{dL}{dx}\right)^2=\tilde f(L)-\tilde f(H),
\end{equation}
which has the implicit solution 
\begin{equation}
  \label{eq:L_x_solution}
  x=\sqrt{\frac{D}{2}} \int\limits_{L(0)}^{L(x)}dL^\prime 
\left[\tilde f(L^\prime)-\tilde f(H)\right]^{-1/2}. 
\end{equation}
Here the origin $x=0$ and corresponding value $L(0) > H$ are arbitrary.  Note
that no boundary conditions need to be specified at large positive $x$,
approaching the meniscus border. This reflects the fact that those boundary
conditions are non-universal and depend on how the film is created
\cite{gem97}.  However, the description based on the above equations is only valid
for small $\nabla_\perp L$; deep in the meniscus, higher-order terms in
$\nabla_\perp L$ should enter \cite{pic01,gem97}.
 
\begin{figure}
\centerline{\epsfxsize=400pt\epsffile{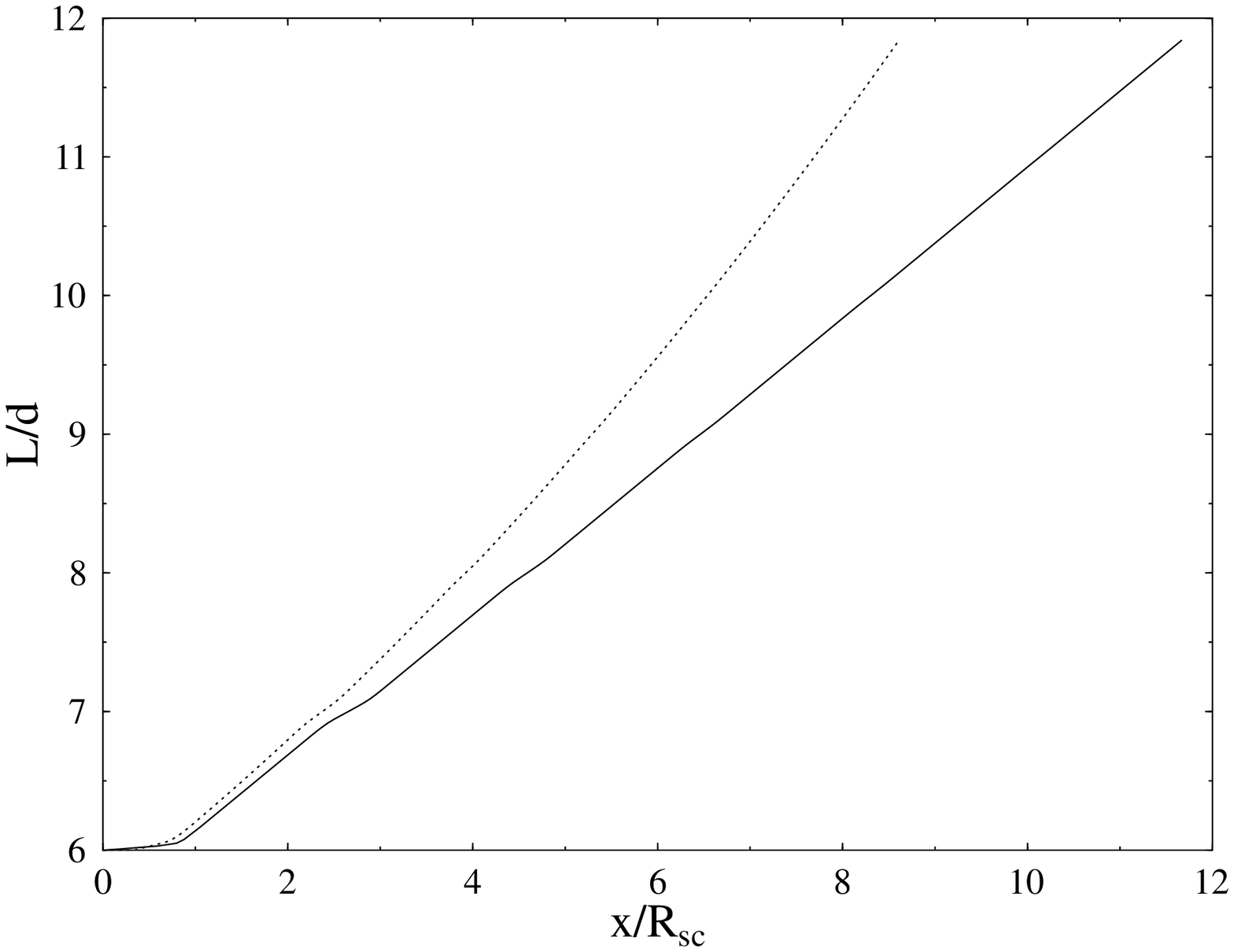}}  
\caption{ 
\label{fig:l_vs_x_rs}
Thickness of the meniscus {\it vs.} scaled in-plane distance $x/R_{sc}$, using the 
$f(L)$ function of Fig.~\ref{fig:f_vs_l_r_s}. The solid line
corresponds to $\Delta P=0$ while the dotted line corresponds to $\Delta
P=0.05C/d^2$. The origin $x=0$ is such that, for both curves, $L(x=0) = 6.00001$  }
\end{figure}

Figure \ref{fig:l_vs_x_rs} shows the solutions to Eq.~(\ref{eq:L_x_solution})
for the $r_s$-model using the same parameters as in Fig.~\ref{fig:f_vs_l_r_s}.
Two cases are shown, the solid line corresponding to $\Delta P=0$ while the
dotted line corresponds to $\Delta P=0.05 C/d^2$. (The reduced unit for
$\Delta P$ is consistent with that for $f$ discussed at the end of
Sec.~\ref{sec:non_intro}. Following from this scaling and
Eq.~(\ref{eq:L_x_solution}), the horizontal distance $x$ is expressed in units
of $R_{sc} \equiv d \sqrt{Dd/C}$.) In both cases, the meniscus profile is
fairly smooth. At low temperatures, when the smectic-A phase is stable in bulk, the
meniscus is expected \cite{pie93} to consist of a set of edge dislocations
which change the film thickness by steps of height $\approx d$. The absence of
such distinct steps in Fig.~\ref{fig:l_vs_x_rs} reflects the fact that here we
are in the overheated regime.

For $T < T_s$, the function $f(L)$ tends to a non-zero value $f(\infty)$ with
increasing $L$. The latter represents the contribution of surface-induced
smectic ordering to twice the interfacial tension $\gamma$ of the liquid-vapor
interface of a semi-infinite liquid. Neglecting small oscillations of $f(L)$
about $f(\infty)$ and defining the meniscus slope angle $2\theta \approx
dL/dx$, Eq.~(\ref{eq:first_integral}) leads to
\begin{equation}
\label{eq:theta}
\theta^2=\frac{1}{2D} \left[ f(\infty)-f(H)+\Delta P(L-H) \right]. 
\end{equation}
On assuming that $D=\gamma/2$ and extrapolating the function in
Eq.~(\ref{eq:theta}) down to $L=H$, we obtain for the contact angle $\theta_m$
between the meniscus and film,
\begin{equation}
\label{eq:theta_m}
\theta_m^2=\frac{1}{\gamma}\left[ f(\infty)-f(H)\right],
\end{equation}
independent of $\Delta P$. This result agrees with that derived in
Ref.~\cite{pic01}; we note that essentially equivalent relations were derived
some time ago in the case of soap films \cite{tos75,def88}. Note also that
Eq.~(\ref{eq:theta}) predicts growth of $\theta$ with increasing $L$. The
meniscus curvature $\kappa$ is found to be
\begin{equation}
\label{eq:R_def}
\kappa \equiv  \frac{d\theta}{dx} =\frac{d \theta}{dL}\frac{dL}{dx}= 
\frac{d\theta^2}{dL}=\frac{\Delta P}{\gamma},  
\end{equation}
which is just the Laplace law. 

\begin{figure}
\centerline{\epsfxsize=400pt\epsffile{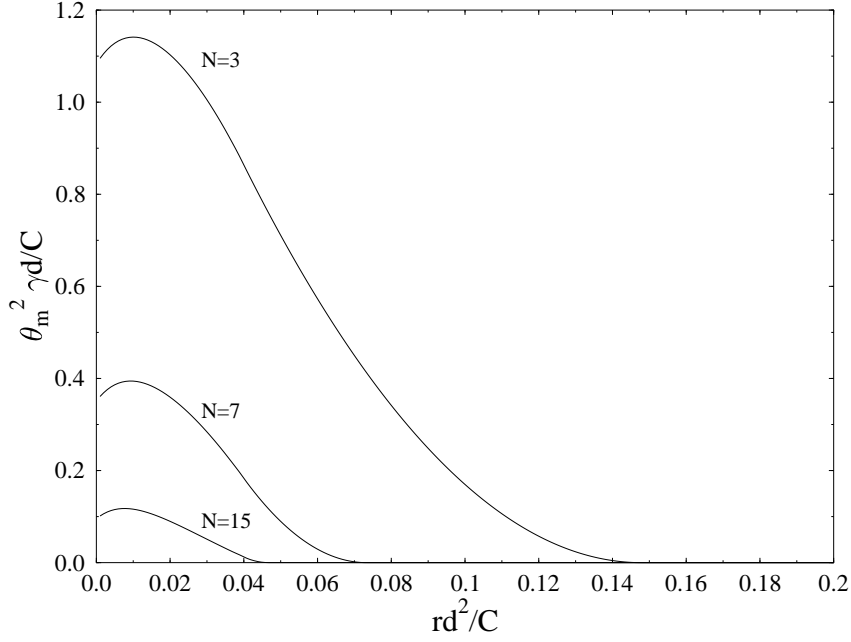}}  
\caption{ 
\label{fig:theta_vs_r}
Squared menisus contact angle $\theta_m^2$ in units of $C/\gamma d$ as a function of
dimensionless temperature variable  $rd^2/C \propto (T-T^*)$ for various number of layers $N$, calculated in the
$r_s$-model with the same parameters as in Fig.~\ref{fig:f_vs_l_r_s}.  }
\end{figure}

Using the original de Gennes model free energy, Eq.~(\ref{eq:f_un}), one finds
that the contact angle diverges as $T$ approaches $T^*$ from above, whereas
the experimental results of Ref.~\cite{pic01} indicate regular behavior in
this region. This feature was alleviated in the calculations of
Ref.\cite{pic01}, based on the de Gennes model, by fixing $\psi(L/2)$ instead
of $h_s$. Although this eliminates the divergence of the contact angle in the
vicinity of the bulk transition point, the resulting model predicts a
different anomaly, namely that $\theta_m$ vanishes as $T \rightarrow T^*$ for
all $N$.  The divergence of the contact angle near the bulk second-order
transition in the original de Gennes model can be removed by setting $g
\neq 0$.  Here we present results for $\theta_m$ using the $r_s$-model,
although qualitatively similar results are obtained using $r_s = 0$ with $h_s
\neq 0$ and $g \neq 0$.  Figure \ref{fig:theta_vs_r} shows $\theta_m^2$, in
units of $C/(d \gamma)$, as a function of $\hat r \propto (T-T^*)$ for various
number of layers $N$, using the free energy $f(L)$ depicted in
Fig.~\ref{fig:f_vs_l_r_s}. It is seen that  $\theta_m$ remains non-zero
on approaching the bulk transition temperature and increases with
decreasing $N$, in agreement with
experiment \cite{pic01}.  We also find that the contact angle
vanishes above the maximum temperature for a given number of layers.  This is
in contrast with the model of Ref.\cite{pic01}, which yields a small but
non-zero contact angle for arbitrarily large temperature and film thickness.
However, as seen in Fig.~\ref{fig:theta_vs_r}, the maximum in $\theta_m$ as a
function of temperature is fairly insensitive to the number of layers, which
does not accord with the experimental results.

Comparison of Fig.~\ref{fig:theta_vs_r} with the experimental data
\cite{pic01} for $\theta_m^2$ indicates that the scaling unit $C/(d\gamma)$
should be on the order of $10^{-2}$. Taking $D=\gamma/2$, we then estimate the in-plane distance
scale unit to be $R_{sc} \equiv d\sqrt{Dd/C} \approx (10/\sqrt{2})d \approx 2 \times 10^{-8}$m,
where we have used the value $d=3 \times 10^{-9}$m \cite{pic00}.

\section{Dynamics of nonuniform films}\label{sec:dynam}
\subsection{Time-dependent Ginzburg-Landau equation}
To study the dynamics of a thinning free-standing smectic film, we 
consider the latter to be a
two-dimensional object with a free energy characterized by
its thickness $L$, given by
Eq.~(\ref{eq:F_approx}) with, generally, $f$ replaced by $\tilde f$. We will focus on dynamical processes with large enough
characteristic times to neglect inertial effects.  
Although the details of
relaxation are undoubtedly quite complicated, here we will proceed by
assuming the simplest possible dissipative dynamics for $L$, based on a time-dependent
Ginzburg-Landau (TDGL) equation \cite{bra94}. This equation is appropriate for
describing the dynamics of a non-conserved variable, which 
$L$ can be regarded in the case of a film open to the exchange of molecules with the
meniscus. The TDGL equation is
\begin{equation}
\label{eq:L_2D_equation}
\eta\frac{\partial L}{\partial t} = -\frac{\delta F}{\delta L} 
= \left[D\nabla^2_\perp L
- \frac{\partial \tilde f}{\partial L} \right].
\end{equation}
Here $\delta F/\delta L$  is the functional derivative of $F$, giving the thermodynamic force which
drives the system toward equilibrium, $t$ is time and $\eta$ is a kinetic
coefficient which we will assume to be constant. 

Two opposing mechanisms for thinning of an overheated smectic film are
``uniform thinning'' (i.e., with $\nabla_\perp L=0$) and via nucleation of dislocation loops.   
If the initial film thickness $L_0 \approx Nd$ is {\it uniform} and at a local
minimum of the shifted free energy $\tilde f$, then it 
will remain so indefinitely according to Eq.~(\ref{eq:L_2D_equation}). 
(This picture neglects possible disruption due to thermal fluctuations, which we
 neglect in this work.) Under small displacements of the thickness from 
the initial value $L_0$, the film will be restored to that initial
thickness. Hence uniform thinning can only occur
at the maximum temperature of an $N$-layer film, and then only  
if $\Delta P > 0$ \cite{foot2}.  

To examine the growth of dislocation loops, for simplicity we  
first consider a one-dimensional solution 
of Eq.~(\ref{eq:L_2D_equation}), in the form of an infinite straight-line  kink
separating domains of thicknesses $L_1$ and $L_2$ and moving along the $x$-axis:
\begin{equation}
  \label{eq:kink_like_solution}
L=\Phi(x-v_\infty t),
\end{equation}
where the function $\Phi$ and kink velocity $v_\infty$ are to be determined. Substituting
Eq.~(\ref{eq:kink_like_solution}) into Eq.~(\ref{eq:L_2D_equation}) yields the ordinary differential equation
\begin{equation}
  \label{eq:Phi_equation}
D\Phi^{\prime\prime}+\eta v_\infty \Phi^\prime -\frac{\partial \tilde
  f(\Phi)}{\partial \Phi}=0~,
\end{equation}
with boundary conditions
\begin{mathletters}
\label{eq:Phi_bound}
\begin{eqnarray}
\label{eq:Phi_bound_a}
\Phi(-\infty)&=&L_1,\\
\label{eq:Phi_bound_b}
\Phi(\infty)&=&L_2,
\end{eqnarray}
\end{mathletters} 
where the prime symbols ($\prime$) denote derivatives of $\Phi$ with 
respect to its argument.
The thicknesses $L_1$ and $L_2$ are at local minima of $\tilde f(L)$. Usually we
will take these to be adjacent minima, with $L_1 \approx (N-1)d$, $L_2
\approx Nd$, and
$\tilde f(L_1) < \tilde f(L_2)$.
    
Equation (\ref{eq:Phi_equation}) has a well-known mechanical analogy
\cite{saa88,but88}. It can be considered as the dynamical equation describing
movement of a particle of ``mass'' $D$ in a medium with ``friction coefficient'' $\eta v_\infty$ 
and subject to a ``potential energy'' $-\tilde f(L)$.
Note that there is no stationary solution ($v_\infty = 0$)
of this equation unless the depths of the minima of $\tilde f(L)$ are equal.
For any function $\tilde f(L)$ characterized by two unequal adjacent
minima, as is the case here, there should be a unique solution for the velocity $v_\infty >0$ 
and function $\Phi$ describing the profile of the kink, which moves
toward the region of thickness $L_2$ in
order to eliminate the domain of higher free energy. Under conditions where 
the higher minimum of $\tilde f(L)$ at $L_2$ vanishes and becomes a point
of zero curvature, as happens in the present model at the maximum temperatures,
the velocity $v_\infty$ and kink shape may become non-unique. In other
contexts \cite{saa88,pop96,sha99}, this is called a state of marginal
stability. Here, we always find (Sec.~\ref{Numres}) that uniform thinning occurs
under these conditions when $\Delta P > 0$ \cite{foot2}.
  
The solutions of the one-dimensional equation Eq.~(\ref{eq:Phi_equation}) 
turn out to be relevant in the more general two-dimensional case,
as discussed in the next subsection.  

\subsection{Nucleation of dislocation loops} \label{Nucleation}
If a dislocation loop separating $N$- and $(N-1)$-layer regions is nucleated, initially it will
be of finite size. According to the conventional nucleation picture
(see, e.g., Refs.~\cite{pan99,gen93,gem97} for the case of smectic films), 
the loop then will either expand or collapse depending on whether its
initial radius is greater or smaller than some ``critical'' value.
Here we are interested in determining the critical radius, the
associated activation free
energy, and the subsequent dynamical evolution of the 
loop. 

We assume that the dislocation
loop is a circle. Using in-plane polar coordinates, with origin at
the center of the loop and radial 
distance denoted $r$,
Eq.~(\ref{eq:L_2D_equation}) becomes
\begin{equation}
  \label{eq:L_2d_polar}
  \eta\frac{\partial L}{\partial t} = \left[D\frac{1}{r}
\frac{\partial}{\partial r}
\left(r \frac{\partial L}{\partial r}\right) 
- \frac{\partial \tilde f}{\partial L} \right]~.
\end{equation}
The associated boundary conditions are
\begin{mathletters}
\label{eq:L_2D_boundary}
\begin{eqnarray}
  \label{eq:L_2D_boundary_a}
\left(\frac{\partial L(r)}{\partial r}\right)_{r=0}&=&0~,\\
  \label{eq:L_2D_boundary_b}
L(\infty) = L_2 &\approx& Nd~.
\end{eqnarray}
\end{mathletters}
The film thickness 
in the center of the loop at $r=0$ will usually be close to the value $L_1 \approx (N-1)d$.  The change
in free energy due to formation of the loop is
\begin{eqnarray}
\label{eq:delta_F_loop}
\nonumber
  \Delta F &=&\frac{1}{2} \int d^2r_\perp 
\left[ 2\tilde f(L)-2\tilde f(L_2)+D(\nabla_\perp L)^2\right] \\
&=&\pi \int\limits_{0}^{\infty}dr r \left[2\tilde f(L)-2\tilde f(L_2)
+D\left(\frac{\partial L}{\partial r}\right)^2\right].
\end{eqnarray}
In the case of a stationary solution of Eq.~(\ref{eq:L_2d_polar}), corresponding
to a ``critical'' nucleus, $\Delta F$ is the activation free energy.
 
The profile of $L(r)$ describing a dislocation loop should have
a kink-like structure, with 
$\partial L/\partial r \approx 0$ everywhere except within
a narrow region centered around some value $r = R$. 
Numerical solution of the stationary limit of Eq.~(\ref{eq:L_2d_polar}),
with $(\partial L/\partial t)=0$ and the boundary conditions
Eq.~(\ref{eq:L_2D_boundary}) is
difficult, precisely because this is associated with a unique
``critical'' value of the loop radius $R$. We have found it more 
expedient, and of more general relevance, to solve the full time-dependent
partial differential equation Eq.~(\ref{eq:L_2d_polar}), using a
standard subroutine (NAG Fortran D03PCF).    
Numerical analysis of the differential equation,
starting from initial trial profiles mimicking the expected kink
structure, shows that
solutions in the form of moving kinks are, indeed, dynamically stable. 
Hence it is worthwhile to consider from the outset a
solution to Eq.~(\ref{eq:L_2d_polar}) in the form of a moving
kink: 
\begin{equation}
  \label{eq:kink_like_solution_polar}
L(r,t) = \Phi\left(r-R(t)\right).
\end{equation}
As discussed some time ago in a general context by Chan \cite{cha77}, this form is not an
exact solution of Eq.~(\ref{eq:L_2d_polar}) but should be a 
good approximation
if the kink radius $R$ is much larger than its width denoted $\Delta R$. 
Using Eq.~(\ref{eq:kink_like_solution_polar}) in Eq.~(\ref{eq:L_2d_polar}) and
approximating $1/r$ by $1/R$, we arrive again at
Eq.~(\ref{eq:Phi_equation}), but with $v_\infty$ defined as \cite{cha77}
\begin{equation}
  \label{eq:v__inf_loop}
v_\infty=\frac{dR}{dt}+ \frac{D}{\eta}\frac{1}{R}. 
\end{equation}
This transforms the two-dimensional problem to the one-dimensional case,
which requires finding the pair
$(\Phi,v_\infty)$ with {\it constant} $v_\infty$, satisfying Eq.~(\ref{eq:Phi_equation}) with boundary
conditions Eqs.~(\ref{eq:Phi_bound}). The quantity $v_\infty$ is 
the asymptotic velocity of a loop with sufficiently large radius $R$ and
depends (for given $D$ and $\eta$) only on the function $\tilde f(L)$ and its chosen pair of minima.
Our numerical analyses
of the original two-dimensional equation Eq.~(\ref{eq:L_2d_polar}) show that
both the kink shape $\Phi$ in the moving coordinate frame and
the right-hand-side of Eq.~(\ref{eq:v__inf_loop}) remain practically constant as the kink moves.
Thus, while the results to be reported in Sec.~\ref{Numres} are all obtained from the numerical
solution of Eq.~(\ref{eq:L_2d_polar}), the one-dimensional mapping based on Eqs.~(\ref{eq:Phi_equation}),
(\ref{eq:kink_like_solution_polar}) and (\ref{eq:v__inf_loop}) usually is an excellent approximation and 
provides, as discussed below, a useful framework for interpreting the results.

Note that setting the time derivative of $R$ equal to zero in Eq.~(\ref{eq:v__inf_loop}) yields an expression for the critical loop
radius, 
\begin{equation}
  \label{eq:R_eq_dyn}
  R_{c}=\frac{D}{\eta v_\infty}.
\end{equation}
If $v_\infty$ is known, then $R(t)$ at arbitrary time can be found by solving
Eq.~(\ref{eq:v__inf_loop}) \cite{cha77}, which gives the relation 
\begin{eqnarray}
  \label{eq:t_vs_R}
\nonumber
  t-t_0&=&\frac{1}{v_\infty}\int_{R_0}^{R} dR^\prime 
\frac{R^\prime}{R^\prime-R_{c}} \\
&=&\frac{1}{v_\infty}
\left(R-R_0+R_{c}\ln\left|\frac{R-R_{c}}{R_0-R_{c}}\right|\right),
\end{eqnarray}
where $R_0$ is the radius at an arbitrary initial time $t_0$.

In conventional treatments of nucleation phenomena,
properties of the critical nucleus such as $R_c$ are expressed
in terms of purely thermodynamic quantities, and it is worthwhile to 
show that connection in the present context. From 
Eq.~(\ref{eq:Phi_equation}), one finds that $\Phi$ and
$v_\infty$ satisfy
\begin{equation}
  \label{eq:full_diff}
\frac{d}{dX}\left[\frac{1}{2}D(\Phi^\prime)^2 - \tilde f(\Phi)\right] =
-\eta v_\infty (\Phi^\prime)^2,
\end{equation}
where $X \equiv x-v_\infty t$ is the argument of $\Phi$. 
Integrating Eq.~(\ref{eq:full_diff}) over $X$ using the boundary
conditions Eq.~(\ref{eq:Phi_bound}) yields
\begin{equation}
  \label{eq:v_vs_E}
  v_\infty=\frac{D}{\eta E}\left[\tilde f(L_2)-\tilde f(L_1)\right],
\end{equation}
where the quantity $E$, which will be identified as the line tension of the
dislocation loop, is defined as
\begin{equation}
\label{eq:E_def}
E \equiv D\int_{-\infty}^{\infty} dX
(\Phi^\prime)^2=D\int_{L_1}^{L_2}d\Phi \Phi^\prime.
\end{equation}
Consistent with the one-dimensional mapping, the lower limit of integration over $X$ has been set
to $-\infty$. To a first approximation, Eq.~(\ref{eq:v_vs_E}) shows that $v_\infty$ is
proportional to the free energy difference $\tilde f(L_2)-\tilde f(L_1)$
and, hence, vanishes if that difference is zero. 
Using Eq.~(\ref{eq:v_vs_E}), Eq.~(\ref{eq:R_eq_dyn}) becomes
\begin{equation}
  \label{eq:R_eq_vs_f}
R_{c}=\frac{E}{\tilde f(L_2)-\tilde f(L_1)}.
\end{equation}
One can show that Eqs.~(\ref{eq:E_def}) and (\ref{eq:R_eq_vs_f}) for
$E$ and $R_c$ 
are consistent with standard arguments of nucleation theory. The free
energy change Eq.~(\ref{eq:delta_F_loop}), with the solution for 
$L$ in the form of
Eq.~(\ref{eq:kink_like_solution_polar}), can be approximated as
\begin{equation}
  \label{eq:Delta_F_estim}
\Delta F =-\pi R^2[\tilde f(L_2)-\tilde f(L_1)]+2\pi RE~.
\end{equation}
This follows on approximating the integrand in Eq.~(\ref{eq:delta_F_loop}) 
by the constant $\tilde f(L_1)-\tilde f(L_2)$ inside a circle of radius $R-\Delta R/2$, neglecting the 
integrand for $r > R+\Delta R/2$, and using
Eq.~(\ref{eq:full_diff}) within the kink region of width $ \Delta R$. 
Equation (\ref{eq:Delta_F_estim}), with $E$ given by Eq.~(\ref{eq:E_def}),  is then
obtained on assuming $\Delta R \ll R$.
The first term in Eq.~(\ref{eq:Delta_F_estim}) is
the decrease in film free energy due to the difference
$\tilde f(L_2)-\tilde f(L_1)$ \cite{foot3} while the second term is the free 
energy which has to be overcome due to the line tension of the loop.
Maximizing
$\Delta F$ with respect to $R$ yields the critical radius $R_{c}$ given by Eq.~(\ref{eq:R_eq_vs_f}). 
The corresponding activation free energy
$F_{act}=\Delta F(R_c)$ is
\begin{equation}
  \label{eq:F_act}
  F_{act}=\frac{\pi E^2}{\tilde f(L_2)-\tilde f(L_1)}=\pi R_{c}E.
\end{equation}

The expression for $E$ in Eq.~(\ref{eq:E_def}) agrees
with a familar mean-field relation for the tension of a stationary interface 
in terms of its profile shape \cite{cha95}. Here that relation
also applies to a moving kink of sufficiently large radius, under 
the assumption (supported by our numerical studies) that the profile shape
is preserved during its motion. 

One final point to note concerns the physical interpretation of the equation of motion Eq.~(\ref{eq:v__inf_loop})
for $R(t)$. That equation is equivalent to the balance of thermodynamic and dissipative forces
per unit length of the dislocation line:
\begin{equation}
  \label{eq:R_vs_t}
\frac{1}{2\pi R}\left(\frac{d\Delta F}{dR}\right)+ \eta \frac{E}{D}\left(\frac{dR}{dt}\right)= 0~. 
\end{equation}
This yields Eq.~(\ref{eq:v__inf_loop}) on using Eqs.~(\ref{eq:v_vs_E}) and (\ref{eq:Delta_F_estim}), and
agrees with the model of dislocation-loop dynamics described by Geminard et al. \cite{gem97}, on
identifying the mobility $\mu$ used by these authors with the quantity $Dd/(\eta E)$. The study of
dislocation-loop dynamics in Ref.~\cite{gem97}, performed {\it below} the bulk A-N 
transition temperature, showed that
the resulting integrated form  Eq.~(\ref{eq:t_vs_R}) very well fits experimental data. 
Performing our own fit of Eq.~(\ref{eq:t_vs_R}) to the data reported in Ref.~\cite{gem97} yields the values $v_\infty=2.59\mu$m/s and $R_{c}=42.6\mu$m,
to be contrasted with results discussed below for overheated films.

\subsection{Numerical results} \label{Numres}
\begin{figure}
\centerline{\epsfxsize=400pt\epsffile{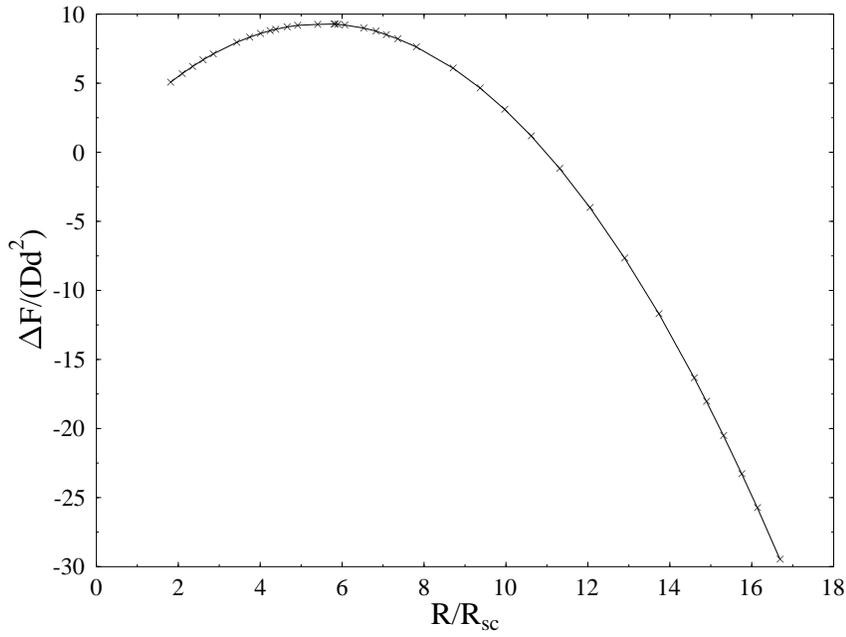}}  
\caption{ 
\label{fig:f_vs_r} 
Free energy $\Delta F$ (reduced by $Dd^2$) of a dislocation loop in a 6-layer film 
{\it vs.} radius $R/R_{sc}$  Crosses: numerical
solution of the dynamical equation Eq.~(\ref{eq:L_2d_polar}) for the same $f(L)$ and 
$r_s$-model parameters as in
Fig.~\ref{fig:f_vs_l_r_s}, with $\Delta P=0$. Solid line: fit of $\Delta F$ with the
parabolic function in Eq.~(\ref{eq:Delta_F_estim}). The best fit yields $E=0.54 \sqrt{DCd}$
and $\tilde f(L_2)-\tilde f(L_1)=0.098 C/d$.  }
\end{figure}

Figure \ref{fig:f_vs_r} illustrates the dependence of the free energy $\Delta F$ on radius $R$ due to a growing dislocation loop 
in a 6-layer film, for the case of $f(L)$ shown in 
Fig.~\ref{fig:f_vs_l_r_s} and $\Delta P=0$.
The in-plane distance
is expressed in units of $R_{sc} \equiv d\sqrt{Dd/C}$, as in Fig.~\ref{fig:l_vs_x_rs},
while the free energy is plotted in units of
$Dd^2$, which follows by scaling of Eq.~(\ref{eq:delta_F_loop}).
The crosses are obtained by monitoring the numerical solution of
Eq.~(\ref{eq:L_2d_polar}) for a dynamically stable kink as it evolves with time, where the loop radius
$R$ is defined such that $L(R)=[L(0)+L(\infty)]/2$. It is seen that $\Delta F(R)$ is very well fit
by Eq.~(\ref{eq:Delta_F_estim}): the fitted value of  $ f(L_2)-
f(L_1)$  agrees with that obtained directly from 
the static theory 
with a precision of 0.5 per cent. We also verified that the 
right-hand-side of Eq.~(\ref{eq:v__inf_loop}) remains constant within
the same precision, giving 
a value $v_\infty=0.18 D/(\eta R_{sc})$. This numerical calculation supports the ansatz
Eq.~(\ref{eq:kink_like_solution_polar}) for a two-dimensional
kink with constant $v_\infty$ related to $R(t)$ by
Eq.~(\ref{eq:v__inf_loop}).  
Analogous fitting to Eqs.~(\ref{eq:Delta_F_estim}) and 
(\ref{eq:v__inf_loop}) of the numerically determined $\Delta F(R)$ and
kink radius has been carried out for all our calculations, showing 
comparable precision except very close to the maximum temperature
of an $N$-layer film, where uniform thinning occurs. As a further
check, we find very close agreement between the values of the line tension $E$ obtained from the fitting to
Eq.~(\ref{eq:Delta_F_estim}) and by direct evaluation of
Eq.~(\ref{eq:E_def}) from the numerically determined kink profiles.   

\begin{figure}
\centerline{\epsfxsize=400pt\epsffile{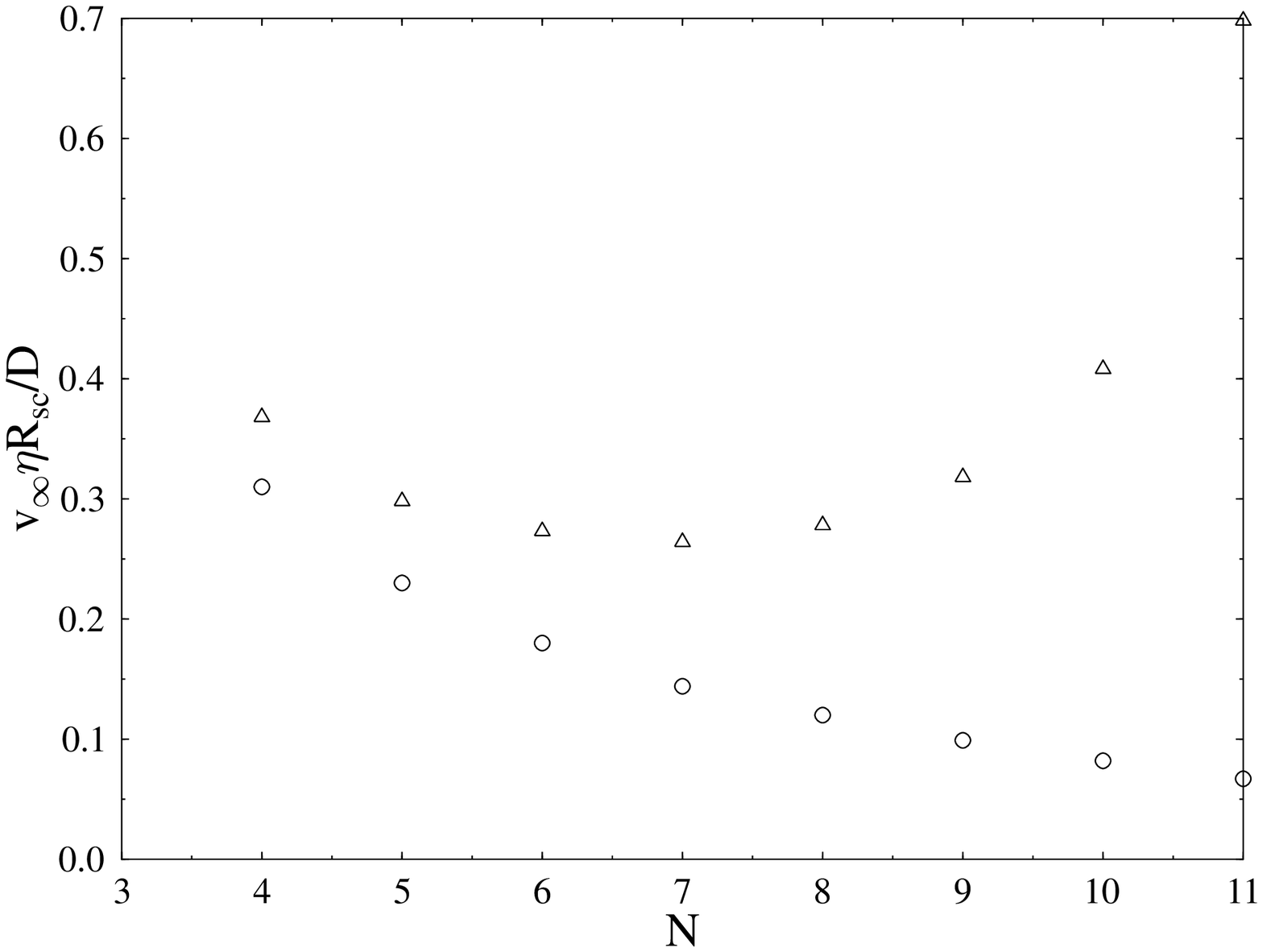}}  
\caption{ 
\label{fig:v_vs_n} 
Velocity $v_\infty$ (reduced by $D/(\eta R_{sc})$) for $N$-layer films
{\it vs.} the number of layers, calculated using the $f(L)$ depicted in
Fig.~\ref{fig:f_vs_l_r_s}. Circles: $\Delta P=0$.  Triangles: $\Delta
P=0.05C/d^2$ (as in Fig.~\ref{fig:l_vs_x_rs}).}
\end{figure}

\begin{figure}
\centerline{\epsfxsize=400pt\epsffile{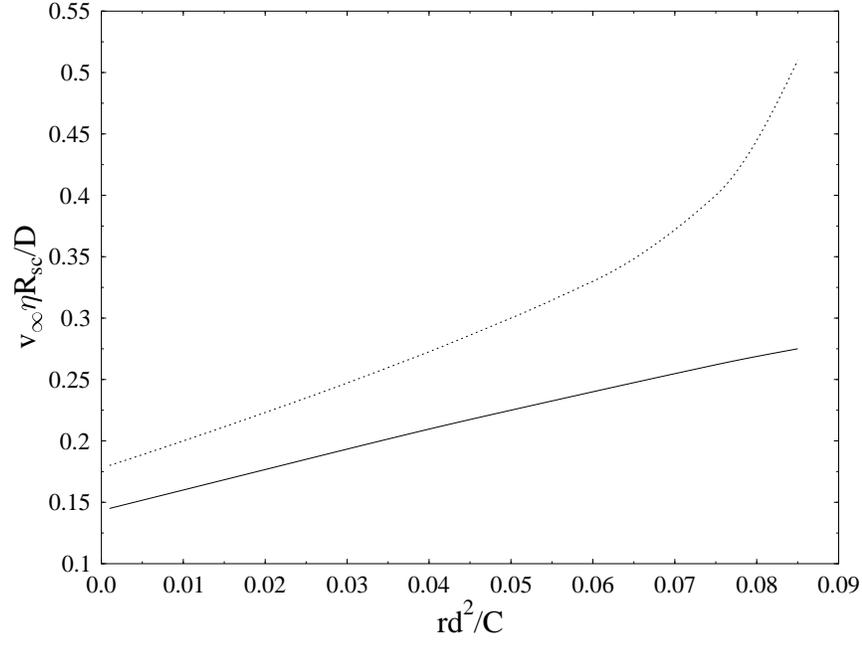}}  
\caption{ 
\label{fig:v_vs_r} 
Reduced velocity for a $5$-layer film
{\it vs.}  $rd^2/C \propto (T-T^*)$. Parameters of the $r_s$-model are the same
as in Fig.~\ref{fig:f_vs_l_r_s}. Lower curve (solid): $\Delta P=0$.  Upper curve (dashed): $\Delta
P=0.05C/d^2$. }
\end{figure}

Only recently, in Ref.~\cite{pan00}, have the dynamics of dislocation loops in overheated films been studied 
experimentally, albeit for a system exhibiting a {\it first-order} bulk A-isotropic (I)
transition. (This difference will be discussed in Sec.~\ref{sec:disc}.) The magnitude of the loop velocity was found to be $10^3$--$10^4$ times larger
than that reported for the smectic-A phase in Ref.~\cite{gem97}, which we
will comment on further below.
We note that the
data in Ref.~\cite{pan00} show a purely linear dependence of $R$ on time, suggesting that
$R/R_{c} \gg 1$ in the experimentally accessible range and that the 
measured velocity
corresponds to $v_\infty$ of the present theory.
The data reported in Ref.~\cite{pan00} also show that the dislocation-loop
velocity slightly increases with increasing number of layers.  To check the dependence of
$v_\infty$ on thickness in our theory we took $\eta$ to be independent of $N$.
Figure \ref{fig:v_vs_n} shows the results
of calculations using the $f(L)$ function employed in Fig.~\ref{fig:f_vs_l_r_s}.
It is seen that in some range of thickness the velocity slightly increases
with $N$ provided that $\Delta P >0$.  We do not rule out that $\eta$
depends on thickness and diminishes with $N$, as suggested in Ref.~\cite{pan00}, but a detailed analysis of 
$\eta$ is beyond the scope of this paper. Figure \ref{fig:v_vs_r} shows, in the
case of a 5-layer film, that $v_\infty$ is predicted to increase with
temperature, in agreement with experiment \cite{pan00}. The rate of
increase is enhanced in the presence of a nonzero $\Delta P$. The upper temperature limits of the curves in
Fig. \ref{fig:v_vs_r} are slightly less than the maximum temperatures for the given values of $N$ and
$\Delta P$. Beyond these limits, dislocation-loop growth is 
superceded by uniform thinning.

Using Eq.~(\ref{eq:R_eq_dyn}) and the velocity data in Fig.~\ref{fig:v_vs_n},
the critical radius $R_c$ can be determined. Referring only to the points 
for $N=11$ in the figure, we find that $R_c=14.9R_{sc}$ and $1.43R_{sc}$
for $\Delta P = 0$ and $\Delta P = 0.05C/d^2$, respectively.
These values bracket the range of $R_c$ values obtained for other values
of $N$. Using the estimate for $R_{sc}$ described at the end of 
Sec.~\ref{sec:non_non}, we thus find $R_c$ to be in the range $10^{-1}$
to $10^{-2}\mu$m, several orders of magnitude smaller than the value deduced above
in the bulk A phase. In view of Eq.~(\ref{eq:R_eq_dyn}), these results
are consistent with the reported differences in loop velocity $v_\infty$
below and above the bulk transition temperature.

\begin{figure}
\centerline{\epsfxsize=400pt\epsffile{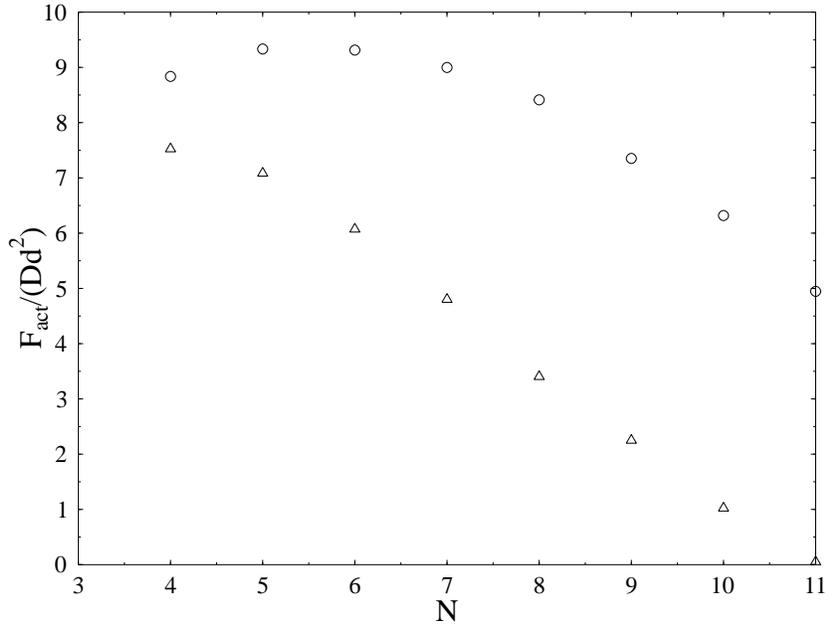}}  
\caption{ 
\label{fig:f_act_vs_n} 
Activation free energy $F_{act}$ (reduced by $Dd^2$) for $N$-layer films {\it vs.}
the number of layers calculated using the $f(L)$ function shown in
Fig.~\ref{fig:f_vs_l_r_s}. Circles: $\Delta P=0$.  Triangles: $\Delta
P=0.05C/d^2$. }
\end{figure}

The key quantity which determines whether spontaneous nucleation of a dislocation loop in an overheated 
film actually occurs is the activation free energy $F_{act}$. 
The dependence of 
$F_{act}$ (in units of $Dd^2$), for a single-layer
dislocation loop,  on the number of layers
$N$ in the initial uniform film is shown in Fig.~\ref{fig:f_act_vs_n}.  Again, the calculations were done 
using the $f(L)$ function
depicted in Fig.~\ref{fig:f_vs_l_r_s}. Note that the temperature in this case is slightly
below the maximum temperature for an 11-layer film. To judge whether nucleation occurs,
we use the argument described in Ref.~\cite{pan99}, based on the frequency per unit area
of forming a dislocation loop of radius $R>R_c$. This frequency is given by  $f=f_0\exp(-F_{act}/k_BT)$,
where $f_0$ is estimated to be $10^{26}$cm$^{-2}$s$^{-1}$ \cite{gen93}.
This yields \cite{pan99} that the condition for a dislocation loop to nucleate
in $0.1$ s in a $1$ cm$^2$ film is $F_{act}/k_BT \approx 60$. To compare this number with 
Fig.~\ref{fig:f_act_vs_n}  requires a value for the dimensionless parameter $Dd^2/(k_BT)$.  
Taking $D=\gamma/2 \approx 1.5\times 10^{-2}$ N/m, $ k_BT
=4.5\times 10^{-21}$J and $d=3\times 10^{-9}$m, we estimate that $Dd^2/k_BT \approx 30$.
These numbers yield the criterion that $F_{act}/(Dd^2)$ must become  about $2$ or smaller for spontaneous
nucleation to occur. Figure \ref{fig:f_act_vs_n} shows that, for the chosen set of
parameters, this can be achieved provided that $\Delta P>0$ is sufficiently large.

\begin{figure}
\centerline{\epsfxsize=400pt\epsffile{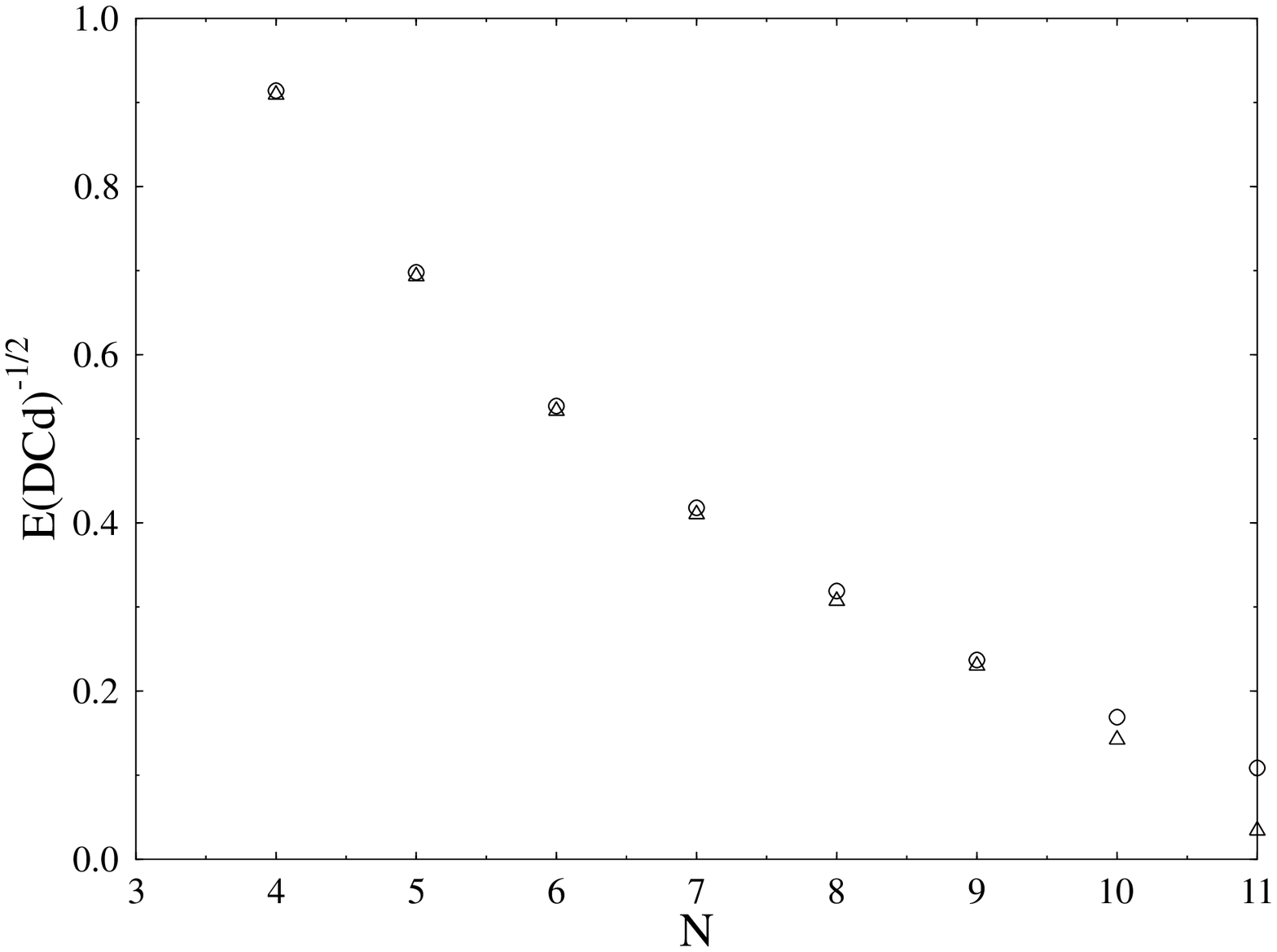}}  
\caption{ 
\label{fig:e_vs_n} 
Line tension $E$ (reduced by $\sqrt{DCd}$) for $N$-layer films {\it vs.} the
number of layers, calculated using $f(L)$ depicted in
Fig.~\ref{fig:f_vs_l_r_s}. Circles: $\Delta P=0$.  Triangles: $\Delta
P=0.05C/d^2$. }
\end{figure}

Figure \ref{fig:e_vs_n} shows the variation of the line tension $E$ with initial number of layers $N$
for the same cases shown in Figs.~\ref{fig:v_vs_n} and \ref{fig:f_act_vs_n}. Except for the last 
two points in the figure ($N=10,11$), $E$ is seen to be essentially independent of $\Delta P$. It 
follows from Eq.~(\ref{eq:F_act}) that it is the decreasing magnitude of $E$ on approaching $N_{cr}$ 
for a given $T$ which is mainly responsible for the decrease in $F_{act}$ in Fig.~\ref{fig:f_act_vs_n}
near $N_{cr}$ and, similarly, for the increase of $v_\infty$ when $\Delta P > 0$ in
Fig.~\ref{fig:v_vs_n}. The difference between the two cases, $\Delta P = 0$ and $\Delta P > 0$, in 
Figs.~\ref{fig:v_vs_n} and \ref{fig:f_act_vs_n} is due to the fact that $\tilde f(L_2) - \tilde f(L_1) \approx
f(L_2)-f(L_1) + \Delta Pd$ is dominated by the term $\Delta Pd$ for the value of the latter used. 

\section{Conclusions}\label{sec:disc} 
 The present theory, a modification of de Gennes' \cite{gen90} theory
of presmectic films, is based on the generally accepted view \cite{pan99} that
the occurrence of overheated free-standing films is due to 
surface-enhanced smectic ordering \cite{bahr94b}. In the case of a uniform planar
film, our theory \cite{sha01} predicts that there is a maximum
temperature  for which smectic ordering in an $N$-layer
film can occur. We associate this with an upper bound for the
true layer-thinning transition temperature $T_c(N)$.
Employing a dynamical generalization of the theory
based on a time-dependent Ginzburg-Landau (TDGL) equation, we have
shown that thinning via nucleation of dislocation loops, the
mechanism indicated by recent experiments \cite{pan99,pan00,pic01},
is possible provided the pressure difference $\Delta P$ resulting
from curvature of the surrounding meniscus is sufficiently large.
Otherwise, the film would undergo either ``uniform thinning'' or,
possibly, rupturing by a process analogous to spinodal dewetting \cite{foot2}. The requirement
for a non-zero $\Delta P$ to promote nucleation of
dislocation loops is consistent with other recent studies
\cite{gor99,pic01}, although we emphasize here that 
the condition $\Delta P \neq 0$ is
not essential for {\it some} type of thinning process to occur.

In the present work, all nucleation properties (i.e., $R_c, 
F_{act}, v_\infty, E$) are interrelated within the framework of 
solutions to the TDGL equation. In particular, the line tension
$E$ of the dislocation loop is expressed in terms of
the kink profile, see Eq.~(\ref{eq:E_def}). From this equation, $E$ depends
implicitly on the elastic behavior of the system through the
dependence of the kink shape on the uniform film free energy
$\tilde f(L)$. 
This self-consistent method for incorporating the line tension differs 
from those used in Refs. \cite{gor99,pan99,pic01,gem97}.

It is important to emphasize that the present work is restricted to
smectic-A liquids undergoing continuous A-N transitions in bulk.
For this reason we do not attempt to make quantitative comparisons  
with the experimental results of Ref. \cite{pan00} for dislocation
loop dynamics, since the latter pertain to a system with a first-order
A-I transition. Such a system can be treated using the present
Landau-de Gennes theory (albeit with considerably 
greater complexity \cite{sha01a}) by appropriately modifying the bulk free-energy density
in Eq.~(\ref{f[eta,u]}). One significant difference which is expected
concerns contributions to the ``$\Delta P$'' term in the resulting 
uniform-film free energy.
In a system with a first-order bulk transition, such a term 
arises even in the absence of meniscus effects, due to the 
grand-canonical \cite{foot1} free-energy difference between a {\it metastable}
bulk smectic-A phase and the isotropic phase \cite{mar97,mar96}.
This effect was recognized in the nucleation theory of 
Ref.~\cite{pan99}, although the latter work otherwise
employed de Gennes' theory for a second-order bulk transition.
Clearly, to compare the present dynamical predictions with experiment, 
it would be of interest to perform measurements on the dynamics of
layer thinning in systems such as that studied in Ref.~\cite{pic01}, which
exhibit second-order bulk A-N transitions.  

Our picture is that dislocation-mediated thinning of an overheated
free-standing smectic film may preempt the uniform-thinning 
mechanism and thus, for given $N$ and fixed values of the model
parameters such as $r_s$, occur at a lower temperature
than predicted \cite{sha01} by considering a purely uniform film.
This is conceivable because {\it all} free-standing film states
are metastable \cite{mar96}. Here we have presented several 
qualitative results in support of this picture. An additional
step would be to evaluate the shift in layer-thinning
transition temperatures $T_c(N)$ from those predicted \cite{sha01}
for a uniform film, and attempt a refitting with experimental data.
This is a somewhat tedious task and is left for future study
\cite{sha01a}. Further studies will also be directed to evaluating
the kinetic coefficient $\eta$ and, as already mentioned, 
extending the present theory to smectic films with first-order bulk transitions.

\acknowledgements 

This study was supported by the Natural Sciences and Engineering Research
Council (Canada).


\begin{thebibliography}{10}

\bibitem{stoprl94}
T. Stoebe, P. Mach, and C.~C. Huang, Phys. Rev. Lett. {\bf 73},  1384  (1994).

\bibitem{joh97}
P.~M. Johnson {\it et~al.}, Phys. Rev. E {\bf 55},  4386  (1997).

\bibitem{pan98}
S. Pankratz, P.~M. Johnson, H.~T. Nguyen, and C.~C. Huang, Phys. Rev. E {\bf
  58},  R2721  (1998).

\bibitem{dem95rc}
E.~I. Demikhov, V.~K. Dolganov, and K.~P. Meletov, Phys. Rev. E {\bf 52},
  R1285  (1995).

\bibitem{mol98}
E.~A.~L. Mol {\it et~al.}, Physica B {\bf 248},  191  (1998).

\bibitem{clu00}
P. Cluzeau {\it et~al.}, Phys. Rev. E {\bf 62},  R5899  (2000). This
work describes observations of layer-thinning transitions in a 
smectic-C liquid crystal. 

\bibitem{bahr94b}
C. Bahr, Int. J. Mod. Phys. {\bf 8},  3051  (1994).

\bibitem{mir96}
L.~V. Mirantsev, Liq. Cryst. {\bf 20},  417  (1996).

\bibitem{mar97}
Y. Martinez-Raton, A.~M. Somoza, L. Mederos, and D.~E. Sullivan, Faraday 
Discuss. {\bf 104}, 111 (1996); Phys. Rev. E {\bf 55},  2030  (1997).

\bibitem{gor99}
E.~E. Gorodetskii, E.~S. Pikina, and V.~E. Podnek, JETP {\bf 88},  35  (1999).

\bibitem{sha01}
A.~N. Shalaginov and D.~E. Sullivan, Phys. Rev. E {\bf 63},  031704  (2001).

\bibitem{pan99}
S. Pankratz, P.~M. Johnson, R. Holyst, and C.~C. Huang, Phys. Rev. E {\bf 60},
  R2456  (1999).

\bibitem{pan00}
S. Pankratz, P.~M. Johnson, A. Paulson, and C.~C. Huang, Phys. Rev. E {\bf 61},
   6689  (2000).

\bibitem{pic01}
F. Picano, P. Oswald, and E. Kats, Phys. Rev. E {\bf 63},  021705  (2001).

\bibitem{gen90}
P.~G. de~Gennes, Langmuir {\bf 6},  1448  (1990).

\bibitem{sul86}
D.~E. Sullivan and M.~M. Telo da Gama, in {\em Fluid Interfacial Phenomena}, 
C.~A. Croxton, ed. (Wiley, 1986) p.45;~ M. Schick, in {\em Les Houches, Session XLVIII, 1988 -
Liquides aux interfaces}, J. Charvolin, J.~F. Joanny and J. Zinn-Justin, eds. (Elsevier, 1990) p.418.

\bibitem{foot1}
As appropriate for an open system, the film free energy in 
Eq.~(\ref{f[eta,u]}) is really the excess grand canonical potential per
unit area: see \cite{mar96}. Note that Eq.~(\ref{f[eta,u]}) only involves
the effects of smectic ordering, and omits the contribution
to the free energy of a background
isotropic density variation across the film, which is assumed to be 
constant.

\bibitem{mar96}
Y. Martinez, A.~M. Somoza, L. Mederos, and D.~E. Sullivan, Phys. Rev. E {\bf
  53},  2466  (1996).

\bibitem{cha95}
P.~M. Chaikin and T.~C. Lubensky, {\em Principles of Condensed Matter Physics} (Cambridge University Press, Cambridge, 1995).

\bibitem{sha01a}
A.~N. Shalaginov and D.~E. Sullivan (to be published).

\bibitem{mor94}
L. Moreau, P. Richetti, and P. Barois, Phys. Rev. Lett. {\bf 73},  3556
  (1994).

\bibitem{gen93}
P.~G. de~Gennes and J. Prost, {\em The Physics of Liquid Crystals} (Clarendon
  Press, Oxford, 1993); J. Prost, Adv. Phys. {\bf 33}, 1 (1984).

\bibitem{fis89}
M.~E. Fisher, in {\em Statistical Mechanics of Membranes and Surfaces},
D. Nelson, T. Piran and S. Weinberg, eds. (World Scientific, Singapore, 1989) p.19. 

\bibitem{tos75}
B.~V. Toshev and I.~B. Ivanov, Col. Pol. Sci. {\bf 253},  558  (1975).

\bibitem{pie93}
P. Pieranski {\it et~al.}, Physica A {\bf 194},  364  (1993).

\bibitem{saf94} 
S.~A. Safran, {\em Statistical Thermodynamics of Surfaces, Interfaces,
and Membranes} (Addison-Wesley, Reading, 1994), Ch.4.

\bibitem{pis00}
L.~M. Pismen and Y. Pomeau, Phys. Rev. E {\bf 62}, 2480 (2000).

\bibitem{gem97}
J.~C. Geminard, R. Holyst, and P. Oswald, Phys. Rev. Lett. {\bf 78},  1924
  (1997).

\bibitem{def88}
J.~A. de Feijter, in {\em Thin Liquid Films, Fundamentals and
Applications}, I.~B. Ivanov, ed (Dekker, New York, 1988), p.1.

\bibitem{pic00}
F. Picano, R. Holyst, and P. Oswald, Phys. Rev. E {\bf 62},  3747  (2000).

\bibitem{bra94}
A.~J. Bray, Adv. Phys. {\bf 43}, 357 (1994).

\bibitem{foot2}
The solution of Eq.~(\ref{eq:L_2D_equation}) at the maximum temperature when $\Delta P=0$ differs
from that for $\Delta P >0$. Since $\partial \tilde f/\partial L =0$ at the initial film thickness
$L_0$, there is no {\it uniform} non-stationary solution of Eq.~(\ref{eq:L_2D_equation}) when $\Delta P =0$.
Numerically-determined solutions for moving kinks (Sec.~\ref{Nucleation}) in this case exhibit
a divergent kink width. A related effect is a long-range decay of the meniscus profile obtained
from Eq.~(\ref{eq:first_integral}). Fluctuations omitted here are probably important in this
situation, e.g., leading to film rupture or thinning by spinodal decomposition.   
 
\bibitem{saa88}
W. van Saarloos, Phys. Rev. A {\bf 37},  211  (1988).

\bibitem{but88}
M. Buttiker and H. Thomas, Phys. Rev. A {\bf 37},  235  (1988).

\bibitem{pop96}
V. Popa-Nita and T.~J. Sluckin, J. Phys. (Paris) II {\bf 6},  873  (1996).

\bibitem{sha99}
A.~N. Shalaginov, L.~D. Hazelwood, and T.~J. Sluckin, Phys. Rev. E {\bf 60},
  4199  (1999).

\bibitem{cha77}
S. Chan, J. Chem. Phys. {\bf 67},  5755  (1977).

\bibitem{foot3}
Note that in other works \cite{gor99,pan99,pic01}, the quantity
$\tilde f(L_2)-\tilde f(L_1)$ has been approximated from 
the outset by $\Delta Pd$. 
This should be valid well in the smectic-A phase, where the minima
of $f(L)$ are expected to be nearly equal, but is not generally true
in the overheated regime.

\end{thebibliography}

\end{document}